\documentclass[12pt]{article}
\usepackage{amssymb,amsmath}
\begin{document}

\title{The Phenomenological Preparation - Registration Arrow of Time and its Semigroup Representation in the RHS Quantum Theory}
\author{Arno R. Bohm \and Raymond Scurek \and \\
Department of Physics\\
The University of Texas \\
Austin, Texas 78712 \\
email: rscurek@physics.utexas.edu}

\maketitle

\begin{abstract}

\footnotesize

The observed probabilities of quantum mechanics possess a time asymmetry which is based on the truism that a state must be prepared before an observable can be measured in it.  While Hilbert space quantum theory cannot incorporate this arrow of time, the Rigged Hilbert Space (RHS) formulation of quantum mechanics provides a theory of time symmetric as well as time asymmetric quantum physics.

\end{abstract}

\section{Introduction}

It is often claimed that quantum mechanics is done in the Hilbert space ${\cal H}$.  However, in practice physicists almost never use the defining (topological) properties of ${\cal H}$, e.g., they rarely discuss issues involving the convergence of infinite sequences of vectors or the topological completeness of their space of states.  Further, physicists routinely consider vectors that do not even belong to ${\cal H}$, such as Dirac's scattering states.  In fact, there are many good reasons to choose something other than the Hilbert space for quantum mechanics.  A better choice is the Rigged Hilbert Space (RHS) $\Phi \subset {\cal H} \subset \Phi^{\times}$, and many arguments supporting this opinion have been presented elsewhere~\cite{RHS.arguments}.  Here, we would like to discuss one in particular.  It will be argued that a quantum mechanical scattering experiment contains an arrow of time which cannot be incorporated into the conventional Hilbert space theory but naturally comes out of the RHS formalism.  Furthermore, this arrow of time shares many characteristics with the radiation arrow of time, which is familiar from classical electrodynamics.

The radiation arrow of time (i.e., the Sommerfeld radiation condition) is well accepted and has even been considered fundamental by some~\cite{Ritz.Ernst}.  As is well known, the radiation arrow of time is a statement of causality expressing the fact that the electromagnetic field at the point ${\vec x}$ and time $t$ is determined by the action of a source a distance $R$ from ${\vec x}$ at the {\it retarded} time $t_r = t - \frac{R}{c} < t$ (and \underline{not} the advanced time $t_a = t + \frac{R}{c} > t)$.\footnote{Maxwell's equations, of course, do not exclude the advanced times.  They are excluded by a choice of boundary conditions~\cite{Ritz.Ernst}.}  In this paper, we argue that a quantum mechanical scattering experiment has an analogous arrow of time.  However, choosing the Hilbert space ${\cal H}$ for our space of states chooses only time symmetric solutions of the Schrodinger equation and, therefore, excludes this arrow of time.  Nevertheless, this arrow of time can be naturally incorporated in a quantum theory using the the Rigged Hilbert Space (RHS).

\section{An arrow of time in scattering phenomena}
\label{sec-arrow.of.time}

A scattering experiment, like every experiment in quantum mechanics, consists of two distinct stages, $^{(1)}$the preparation of the state and $^{(2)}$the detection of an observable in that state.  It is an obvious statement of causality that {\it a state needs to be prepared before an observable can be measured in it}.  We call this truism the preparation $\Rightarrow$ registration arrow of time~\cite{Ludwig}.  In particular, when applied to resonance scattering, this arrow of time is just the trivial statement that the decay products cannot be detected before the decaying state is produced.

Let $t_0$ be the time at which a state $W$ is prepared\footnote{$W$ is a density operator that may denote a mixed state or, if $W = |\phi\rangle\langle\phi|$, a pure state.} and let $\Lambda$ be an observable which one tries to detect in $W$.  The probability ${\cal P}_W(\Lambda(t))$ to find the observable $\Lambda$ in the state $W$ at a time $t$ is theoretically given by ${\cal P}_W^{theo}(\Lambda(t)) \equiv Tr(W(t)\Lambda) = Tr(W\Lambda(t))$ but experimentally measured as a ratio of detector counts

\begin{equation}
{\cal P}_W^{exp}(\Lambda(t)) \equiv \frac{N(t)}{N} \,\,\,\,\,\, \mbox{for} \,\,\,\,\,\, t > t_0.
\end{equation}

\noindent $N(t)$ is the number of times the observable $\Lambda$ is detected in the state $W$ (i.e., the number of counts) while $N$ is the total number of states $W$ that are produced.  $N(t)$ increases with time and $\frac{N(t)}{N}$ can only take non-negative rational values less than or equal to $1$.

Before the state $W$ has been prepared, the experimental probability must satisfy

\begin{equation}
{\cal P}_W^{exp}(\Lambda(t)) = 0 \,\,\,\,\,\, \mbox{for} \,\,\,\,\,\, t < t_0.     \label{eq: experimentalfact}
\end{equation}

\noindent If there are any counts before $t=t_0$, they would be interpreted as noise and excluded from ${\cal P}_W^{exp}(\Lambda(t))$.  This fact is nothing more than a statement of our preparation $\Rightarrow$ registration arrow of time and is similar to the radiation arrow of time in that one cannot detect the electromagnetic field before it has been emitted by the source.  Consequently, one would like a theory which incorporates this arrow of time by predicting the time evolution of some states for $t \geq t_0$ only.  Then, we would have

\begin{equation}
{\cal P}^{theo}_W(\Lambda(t)) \equiv Tr(W(t)\Lambda) = 0 \,\,\,\,\,\, \mbox{for} \,\,\,\,\,\, t < t_0.  \label{eq: desiredproperty}
\end{equation}

\noindent However, this is impossible in the Hilbert space ${\cal H}$ as is explained below.

\section{The Hilbert space}

In this section we will do little more than give the definition of a Hilbert space, state a few theorems and explain their significance to our present purpose.

Let $\Phi_{alg}$ be a linear scalar product space (with no topology) which contains the relevant states for our physical system.  One obtains the Hilbert space ${\cal H}$ by completing $\Phi_{alg}$ with respect to the norm.  In other words, one adjoins to $\Phi_{alg}$ all limit elements of Cauchy sequences that converge with respect to the norm topology.  We denote this topology by $\tau_{\cal H}$.  In ${\cal H}$, one has the following theorems:

\noindent {\bf Theorem 1} (Gleason) Let $\Lambda$ denote the elements of the set of projection operators.  Then, for every function ${\cal P}(\Lambda)$ that fulfills the requirements of a probability\footnote{${\cal P}(\sum_n \Lambda_n) = \sum_n {\cal P}(\Lambda_n)$ and ${\cal P}(\Lambda) = 1$}, there exists a positive trace class operator $\rho$ such that ${\cal P}(\Lambda) = Tr(\Lambda \rho)$~\cite{Gleason}.

\noindent {\bf Theorem 2} (Stone-von Neumann) The solutions of the Schroedinger-von Neumann equations for the above $\rho$ are time symmetric and given by the group $U(t) = e^{iHt}$ of unitary operators~\cite{Stone-von.Neumann}, i.e., 

\begin{equation}
\rho(t) = U^{\dagger}(t) \rho(0) U(t)
\end{equation}

\noindent in the Schrodinger picture or

\begin{equation}
\Lambda(t) = U(t) \Lambda(0) U^{\dagger}(t)
\end{equation}

\noindent in the Heisenberg picture.

\noindent {\bf Theorem 3} (Hegerfeldt) For every self-adjoint and semi-bounded Hamiltonian $H$, one has either

\begin{equation}
Tr(\Lambda(t) \rho) = Tr(\Lambda \rho(t)) = 0 \,\,\,\,\,\, \mbox{for} \,\,\,\,\,\, -\infty < t < \infty
\end{equation}

\noindent or

\begin{eqnarray}
Tr(\Lambda(t) \rho) = Tr(\Lambda \rho(t)) > 0 \,\,\,\,\,\, \mbox{for} \,\,\,\,\,\, -\infty < t < \infty   \nonumber  \\
	 \mbox{except on a set of Lesbesgue measure zero~\cite{Hegerfeldt}.}
\end{eqnarray}

Theorem 1 states that any probability can be given by a trace.  Then, it follows from Theorem 3 that there is no state in ${\cal H}$ that can have our desired property (\ref{eq: desiredproperty}) without also having a probability that is trivially zero for all $t$ (except on a set of Lesbesgue measure zero).  Consequently, ${\cal H}$ cannot accommodate the preparation $\Rightarrow$ registration arrow of time.

\section{The Rigged Hilbert Space}

Theorem 2 of the preceding section states that time evolution in ${\cal H}$ is given by a unitary group $U(t) = e^{iHt}$ for $-\infty < t < \infty$.  However, in light of (\ref{eq: desiredproperty}), it may be more appropriate to postulate that time evolution should be given by a semigroup, i.e., if we choose the time of preparation to be $t_0 = 0$ it would be something like

\begin{equation}
U^{\dagger}(t) = \mbox{``}\,e^{-iHt}\,\mbox{''} \,\,\,\,\,\, \mbox{for} \,\,\,\,\,\, 0 \leq t < \infty \,\,\,\mbox{only}.    \label{eq: semigroup}
\end{equation}

\noindent Although such a time evolution semigroup cannot exist in the Hilbert space ${\cal H}$, it does have a natural place in the RHS formalism.

The RHS or Gelfand Triplet is a triplet of spaces defined by three different topological completions of the same algebraic (linear scalar product) space $\Phi_{alg}$:

\begin{equation}
\Phi \subset {\cal H} \subset \Phi^{\times}.   \label{eq: RHS}
\end{equation}

\noindent ${\cal H}$ is the usual Hilbert space completed with respect to the topology of the norm.  ${\cal H}$ is not of primary interest to us with regards to the physics, but it is a mathematical necessity in the Rigged Hilbert Space.  Hilbert space mathematics is well established and because of (\ref{eq: RHS}) one can often make statements about $\Phi$ and $\Phi^{\times}$ simply because one knows a lot about ${\cal H}$, and one can define needed terms with reference to ${\cal H}$.  In contrast, the space $\Phi$ is a completion with respect to a countable number of norms, one of which is the Hilbert space norm.  These norms are chosen in order to make the algebra of observables continuous when acting on $\Phi$.\footnote{In general, the algebra of observables is not continuous when acting on ${\cal H}$.  For example, even in the simple case of the one-dimensional harmonic oscillator the position and momentum (or, equivalently, the creation and annihilation operators) fail to be continuous on ${\cal H}$~\cite{Bohm.Gadella}.}  We call this topology $\tau_{\Phi}$.  Since the $\tau_\Phi$-completion of $\Phi_{alg}$ involves adjoining the limit elements of all sequences that are Cauchy sequences with respect to a countable number of norms (including the Hilbert space norm), the requirements on a $\tau_{\Phi}$-Cauchy sequence are more stringent than the requirements on a $\tau_{\cal H}$-Cauchy sequence.  Therefore, there are fewer $\tau_{\Phi}$-Cauchy sequences, and $\Phi \subset {\cal H}$.  The space $\Phi^{\times}$ is the space of continuous antilinear functionals $|F\rangle$\footnote{We will use bras $\langle \,\cdot\, |$ and kets $| \,\cdot\, \rangle$ to denote generalized vectors in $\Phi^{\times}$ and the symbols $(\,\cdot\,|$ and $|\,\cdot\,)$ to denote vectors in $\Phi$ or ${\cal H}$.  Of course, every $|\,\cdot\,)$ and $( \,\cdot\, |$ can also be written as a ket or bra, respectively, because of (\ref{eq: RHS}).} on $\Phi$:

\begin{equation}
|F\rangle : \phi \rightarrow F(\phi) = (\phi | F\rangle \in \mathbb{C} \,\,\, \forall \phi \in \Phi
\end{equation}

\noindent The space of continuous antilinear functionals on ${\cal H}$ is ${\cal H}$, and it can be shown that ${\cal H} \subset \Phi^{\times}$, giving the result (\ref{eq: RHS}).\footnote{More precisely, the space of antilinear functionals on ${\cal H}$, denoted by ${\cal H}^{\times}$, can be shown to be isomorphic to ${\cal H}$ and then identified with ${\cal H}$ giving ${\cal H} = {\cal H}^{\times}$.  It can also be shown that ${\cal H}^{\times} \subset \Phi^{\times}$.  Then, after identifying ${\cal H}^{\times}$ with ${\cal H}$, one has $\Phi \subset {\cal H} = {\cal H}^{\times} \subset \Phi^{\times}$.}

For every {\it $\tau_{\Phi}$-continuous} linear operator $A$, we define its dual or conjugate operator $A^{\times}$ on the space $\Phi^{\times}$ by

\begin{equation}
(A\phi | F\rangle \equiv (\phi | A^{\times} | F\rangle \,\,\, \forall \phi \in \Phi \,\,\, \mbox{and} \,\,\, \forall F \in \Phi^{\times}.  \label{eq: A.conjugate}
\end{equation}

\noindent Since $A$ is a $\tau_{\Phi}$-continuous operator and $|F\rangle$ is a continuous functional, $A^{\times}$ is a continuous linear operator on $\Phi^{\times}$.  For any observable $A$ ($\tau_{\Phi}$-continuous), one therefore has a triplet of operators

\begin{equation}
A^{\dagger}|_{\Phi} \subset A^{\dagger} \subset A^{\times}
\end{equation}

\noindent where $A^{\dagger}$ is the Hilbert space adjoint of (the closure of) $A$ and $A^{\dagger}|_{\Phi}$ is its restriction to the subspace $\Phi$.  Equation (\ref{eq: A.conjugate}) gives $A^{\times}$ as the (uniquely defined) extension of $A^{\dagger}$ to the space $\Phi^{\times}$.

One can now define generalized eigenvectors of $\tau_{\Phi}$-continuous operators.  $|F\rangle \in \Phi^{\times}$ is a {\it generalized eigenvector} of the $\tau_{\Phi}$-continuous operator $A$ iff for some complex number $\omega$ one has

\begin{equation}
(A\phi | F\rangle = (\phi | A^{\times} | F\rangle = \omega (\phi | F\rangle \,\,\, \forall \phi \in \Phi.      \label{eq: geneigenvalue}
\end{equation}

\noindent Since (\ref{eq: geneigenvalue}) is valid for every $\phi \in \Phi$, it is often written as a functional equation on $\Phi$:

\begin{equation}
A^{\times}|F\rangle = \omega |F\rangle
\end{equation}

\noindent The eigenvalue $\omega$ may be part of a discrete or a continuous spectrum.

If $H$ is the (essentially) self-adjoint Hamiltonian for a particular scattering system, then $\Phi^{\times}$ will contain generalized eigenvectors of $H$ which have the properties of Dirac's scattering states:

\begin{equation}
H^{\times}|E\rangle = E|E\rangle, \,\,\,\,\, E \geq 0.
\end{equation}

\noindent Further, $\Phi^{\times}$ can also contain generalized eigenvectors of $H$ with complex eigenvalues.  These eigenvalues do not belong to the Hilbert space spectrum of $H$.

\begin{equation}
H^{\times}|E_R - i\frac{\Gamma}{2}\rangle = (E_R - i\frac{\Gamma}{2})|E_R - i\frac{\Gamma}{2}\rangle   \label{eq: complex.eigenvalue}
\end{equation}

\noindent These generalized eigenvectors are associated with the second sheet pole of the S-matrix at the complex energy $E-i\frac{\Gamma}{2}$ and represent decaying states.  They are called Gamow vectors.

For empirical reasons~\cite{Bohm.et.al} which shall not be discussed in this short paper, it is conjectured that scattering phenomena are described by a pair of Rigged Hilbert Spaces.  There is a RHS corresponding to the prepared states or in-states $\phi^+ \in \Phi_-$,

\begin{equation}
\Phi_- \subset {\cal H} \subset \Phi_-^{\times},  \label{eq: in.states}
\end{equation}

\noindent and one corresponding to the registered observables $|\psi^-\rangle\langle\psi^-|$ or out-states $\psi^- \in \Phi_+$,

\begin{equation}
\Phi_+ \subset {\cal H} \subset \Phi_+^{\times}.  \label{eq: out.observables}
\end{equation}

\noindent The apparent schizophrenic notation is adopted in order to follow two separate conventions.  In-~(out-)~states in scattering theory conventionally carry a $+$ ($-$) superscript corresponding to the $+i\epsilon$ ($-i\epsilon$) in the Lippman-Schwinger equation.  On the other hand, the energy representation of $\Phi_-$ ($\Phi_+$) is the restriction to the positive real line of the space of well-behaved Hardy functions in the lower (upper) half of the complex energy plane.~\cite{Baumgartel}  So, the vectors carry the usual convention used by physicists, while the spaces carry the usual convention used by mathematicians.

The space $\Phi_-^{\times}$ ($\Phi_+^{\times}$) contains Dirac's in-~(out-) scattering states\footnote{Dirac's scattering states $|E^{\pm}\rangle$ are not in the spaces of physical states $\Phi_{\mp}$.  From a physical point of view, this is because these states cannot be prepared or registered in a scattering experiment.  In the position representation these are the plane waves.  If such a state could be prepared, it could not be localized in time and would only evolve through an inconsequential phase factor.  The best that can be produced or registered is a wave packet, and wave packets are contained in ${\Phi_{\mp}}$.}

\begin{equation}
H^{\times}|E^{\pm}\rangle = E|E^{\pm}\rangle, \,\,\,\,\,\,\,\, E \geq 0 \,\,\, \mbox{and} \,\,\, |E^{\pm}\rangle \in \Phi_{\mp}^{\times}
\end{equation}

\noindent and Gamow vectors

\begin{equation}
H^{\times}|E \pm i\frac{\Gamma}{2}^{\pm}\rangle = (E \pm i\frac{\Gamma}{2})|E \pm i\frac{\Gamma}{2}^{\pm}\rangle, \,\,\,\,\,\,\,\, |E \pm i\frac{\Gamma}{2}^{\pm}\rangle \in \Phi_{\mp}^{\times}.  \label{eq: Gamow.vector}
\end{equation}

\noindent It can be shown that the intersection of $\Phi_+$ and $\Phi_-$ is not empty.  In fact, $\Phi_+ \cap \Phi_-$ is dense in ${\cal H}$.  With this interpretation of (\ref{eq: in.states}) as the space of prepared states and of (\ref{eq: out.observables}) as the space of registered observables, we can choose the time $t=t_0$ of section \ref{sec-arrow.of.time} as $t=0$.~\cite{Bohm.et.al}

An entirely unforeseen result \cite{semigroup.discovery} was that the conjugate operators $U^{\times}_{\pm}(t)$ of the time evolution operators $U(t)|_{\Phi_{\pm}} = e^{iHt}|_{\Phi_{\pm}}$ cannot be defined for all values of time $-\infty < t < \infty$ but only for $t \leq 0$ in the RHS (\ref{eq: in.states}) and only for $t \geq 0$ in the RHS (\ref{eq: out.observables}).  Consequently, time evolution in each RHS is given by a semigroup.  The semigroup which evolves elements of $\Phi_+^{\times}$ is

\begin{equation}
U_+^{\times}(t) = e_+^{-iH^{\times}t} \,\,\,\, \mbox{for} \,\, 0 \leq t < \infty \,\, \mbox{only}.   \label{eq: outsemigroup}
\end{equation}

\noindent This is the precise statement of equation (\ref{eq: semigroup}), which was postulated earlier on physical grounds.

In contrast, the semigroup which evolves elements of $\Phi_-^{\times}$ is

\begin{equation}
U_-^{\times}(t) = e_-^{-iH^{\times}t} \,\,\,\, \mbox{for} \,\, -\infty < t \leq 0 \,\, \mbox{only}.    \label{eq: insemigroup}
\end{equation}

The preparation $\Rightarrow$ registration arrow of section \ref{sec-arrow.of.time} is expressed by (\ref{eq: outsemigroup}) and thus naturally follows from the construction of the RHS of Hardy class energy wavefunctions for scattering phenomena.

\section{Conclusion}

The Rigged Hilbert Space was first introduced into quantum mechanics to give a mathematical justification to Dirac's heuristic bra-ket formalism.  It was only much later \cite{Bohm.QM.book} that it was discovered that the Rigged Hilbert Space can also contain kets like (\ref{eq: Gamow.vector}) which were envisioned by Gamow \cite{Gamow} and that time evolution for scattering phenomena is given by two semigroups.  Then, it was realized that these semigroups represent an ``arrow of time'' akin to the one found in classical electrodynamics.  Indeed, it would be very strange if the radiation arrow of time existed classically but had no quantum counterpart.  Therefore, it seems quite natural that there exists an arrow of time which is an extension of the classical radiation arrow to quantum mechanics.

This arrow of time does not exist in the usual Hilbert space quantum mechanics where time evolution is given by a unitary group and probabilities cannot be zero for any finite time interval without being trivially zero for all times.  However,  the Rigged Hilbert Space naturally incorporates this arrow of time as is expressed by the semigroup time evolution of equations (\ref{eq: outsemigroup}) and (\ref{eq: insemigroup}).

\section*{Acknowledgments}

We would like to thank H.~D.~Doebner for his hospitality at Goslar and Clausthal, and we gratefully acknowledge the support of the Welch Foundation.


\begin{thebibliography}{99}

\bibitem{RHS.arguments} A.~Bohm, S.~Maxson, M.~Loewe and M.~Gadella, Physica {\bf A236} (1997) 485; A.~Bohm and N.~L.~Harshman, {\it Irreversibility and Causality: Semigroups and Rigged Hilbert Spaces}, Springer, Berlin (1998) 181; A.~Bohm, M.~Gadella and S.~Maxson, Computers Math.~Applic.~{\bf 34}, (1997) 427.

\bibitem{Ritz.Ernst} R.~Ritz, Physikalische Zeitschrift 9 (1908) 303; R.~Ritz, Physikalische Zeitschrift 10 (1909) 224.

\bibitem{Ludwig} G.~Ludwig, {\it Foundations of Quantum Mechanics}, Vol.~I, Springer-Verlag, Berlin (1983) and Vol.~II (1985); {\it An Axiomatic Basis of Quantum Mechanics}, Vol.~I, Springer-Verlag, Berlin (1983) and Vol.~II (1987); K.~Kraus, {\it State, Effects, and Operations}, Springer Lecture Notes in Physics 190, Springer-Verlag, Berlin (1983).

\bibitem{Gleason} A.~M.~Gleason, J.~Math.~{\bf 6}, 885 (1957).

\bibitem{Stone-von.Neumann} M.~H.~Stone, Ann.~of Math.~{\bf 33}, 643 (1932).

\bibitem{Hegerfeldt} G.~C.~Hegerfeldt, Phys.~Rev.~Lett.~{\bf 72}, 596 (1994).

\bibitem{Bohm.Gadella} A.~Bohm and M.~Gadella, {\it Dirac Kets, Gamow Vectors and Gel'fand Triplets}, Lecture Notes in Physics {\bf 348}, Springer-Verlag, Berlin (1989).

\bibitem{Bohm.et.al} A.~Bohm, I.~Antoniou, P.~Kielanowski, J.~Math.~Phys.~{\bf 36}, 2593 (1995).

\bibitem{Baumgartel} Originally the Hardy class property was introduced in order to derive (\ref{eq: complex.eigenvalue}) as a generalized eigenvalue equation and to obtain a Breit-Wigner energy distribution for the vector associated to the resonance pole of the S-matrix.  H.~Baumgartel mentioned the Hardy class functions in a private communication.  See also H.~Baumgartel, Math.~Nachr.~{\bf 75} (1976) 133.

\bibitem{semigroup.discovery} A.~Bohm, J.~Math.~Phys.~{\bf 22} (1981) 2813; Lett.~Math.~Phys.~{\bf 3} (1978) 455.

\bibitem{Bohm.QM.book} A.~Bohm, {\it Quantum Mechanics}, 1$^{st}$ Ed.~, Springer, New York (1979); see also 3$^{rd}$ Ed.~ 3$^{rd}$ revised printing, Ch.~XXI (1999, to appear).

\bibitem{Gamow} G.~Gamow, Z.~Phys.~{\bf 51} (1928) 204.

\end{thebibliography}
\end{document}